# Three-Phase Evaluation of

# AI-Assisted Software Development Life Cycle

*Dr. Joshua Strübel, Professor Carrie Russell, Carson Crockett,*
*Jason Ferraro, Nathan Londhe, Uzayr Syed, and Jacob Viehe*

***Abstract* - This paper presents an exploratory evaluation of how increasing levels of AI autonomy affect software development productivity, requirement adherence, and developer cognitive workload. A team of four developers reimplemented the same full-stack web application across three sequential phases: partial AI-assisted development using GitHub Copilot, an AI-exclusive workflow using GitHub Copilot, and an AI-exclusive workflow using AWS Kiro. Evaluation metrics included development effort (hours), requirement adherence (RITM score), AI-interaction efficiency, and NASA-TLX workload measures.**

**Across phases, higher levels of AI autonomy were associated with reduced development effort, improved requirement adherence, and lower self-reported mental workload, while developer frustration increased modestly. The AWS Kiro phase achieved the strongest overall performance on most measured dimensions, suggesting that tooling architecture may influence outcomes independently of AI autonomy level.**

***Index Terms—AI-assisted software engineering, agentic AI, software productivity, developer cognition, human-AI collaboration, GitHub Copilot, AWS Kiro***

## I. INTRODUCTION

The proliferation of AI-powered coding assistants has substantially changed how software engineers approach software development. Tools such as GitHub Copilot and Amazon Kiro have evolved from supplemental systems into autonomous agents capable of planning, implementing, and iterating on entire features with minimal human intervention [7]. GitHub Copilot operates as an in-editor assistant that generates code suggestions, and in its agentic mode, can autonomously implement multi-step features from natural language prompts, and detect and resolve bugs without specific instruction. Amazon Kiro, a spec-driven development tool, takes a more structured approach: developers author formal specification documents, and the agent decomposes and implements features with minimal iterative prompting. Together, these platforms represent meaningfully different architectural philosophies for human-AI collaboration in software development. As adoption accelerates, organizations face a practical question: beyond simple code completion, how do different levels of AI autonomy and agentic platforms affect measurable development outcomes?

Prior controlled research has established that AI-assisted development can yield meaningful productivity gains. Peng et al. found that developers using GitHub Copilot completed an HTTP server implementation 55.8% faster than a control group [1], and larger field experiments have corroborated these results across diverse engineering contexts [2]. However, most existing work focuses on partial AI assistance rather than fully agentic scenarios in which developers serve primarily as orchestrators rather than implementers [7]. Published controlled empirical comparisons of competing agentic platforms under matched conditions remain limited at the time of writing.

This study addresses these gaps through a three-phase controlled evaluation in which the same development team reimplemented the same application with systematically
varied AI assistance levels and tooling. The primary contributions are:

- A controlled, multi-phase empirical comparison of partial and fully agentic AI-assisted development with respect to productivity, quality, and cognitive load dimensions.
- A quantitative comparison of two commercially available agentic platforms (GitHub Copilot and AWS Kiro) under equivalent task conditions.
- A replicable evaluation framework for future comparative studies of AI coding tools.

## II. RELATED WORK

### *AI-Assisted Software Development*

Recent research has demonstrated that AI-assisted programming tools can significantly improve developer productivity and accelerate software delivery [1, 2]. At the same time, researchers have noted important limitations, including inconsistent code reliability, validation overhead, and potential increases in technical debt when AI-generated implementations are insufficiently reviewed [3–5]. Prior work has also highlighted the growing importance of prompt engineering and human oversight in effectively integrating AI assistants into software engineering workflows [6].

### *Agentic Development Workflows*

More recent studies have explored agentic AI development environments in which AI systems autonomously decompose tasks, generate implementations, and iteratively refine software artifacts with limited direct coding by developers [7–9]. These workflows shift the developer role from manual implementation toward supervision, orchestration, and validation of AI-generated outputs. Empirical evaluations of competing agentic platforms under controlled conditions remain limited.

### *Cognitive Load and Human Oversight*

The NASA Task Load Index (NASA-TLX) is a widely validated instrument for measuring perceived workload across dimensions including mental demand, effort, and frustration [10]. Prior software engineering research has used NASA-TLX to evaluate how automation influences developer experience and task complexity considerations [10]. This metric is especially relevant in agentic environments where developers spend less time writing code and more time validating and supervising AI-generated implementations.

### *Research Gap*

Controlled comparative studies evaluating multiple agentic development workflows using matched tasks, software quality metrics, and cognitive workload assessments remain scarce. This study addresses that gap by systematically comparing traditional AI-assisted coding with a fully agentic AI development environment across productivity, quality, and workload dimensions.

## III. RESEARCH QUESTIONS

This study is organized around four primary research questions:

**RQ1:** How does increasing AI autonomy affect software development productivity (development hours)?

**RQ2:** How does increasing AI autonomy affect code quality (RITM score)?

**RQ3:** How does increasing AI autonomy affect developer cognitive load (NASA-TLX)?

**RQ4:** When AI autonomy is held constant, does platform choice (GitHub Copilot vs. AWS Kiro) affect outcomes?

## IV. METHODOLOGY

### *Study Design*

This study was conducted across three sequential phases, each involving an independent reimplementation of the same software product by a team of four student developers. The intended experimental variable was the level and tool of AI-coding assistance; other variables including developer familiarity with the application, requirements, and agentic workflows accumulated across phases represent a primary confound. AI-coding assistance is defined as the percentage of code directly generated by a large language model (LLM); any line written or subsequently modified by a developer is classified as human-written.

Phase 1 (Partial Agentic, GitHub Copilot) represented a modern AI-forward environment: developers used Copilot as the primary code generation method but retained the ability to write and modify code manually.

Phase 2 (AI-Exclusive, GitHub Copilot) implemented an AI-exclusive workflow. Developers orchestrated and validated AI-generated outputs via structured prompt sequences directed at Copilot's agentic mode, minimizing manual code authorship. Human-written lines consisted primarily of corrections and integrations.

Phase 3 (AI-Exclusive, Amazon Kiro) applied the same AI-exclusive workflow using AWS Kiro's specification-driven architecture. Developers produced structured tasks and specification documents; Kiro then planned and generated full feature implementations autonomously, reducing iterative prompt refinement.

### *Participants*

The team consisted of four senior Clemson University computer science students, each maintaining a consistent role across all phases. All had prior Python and JavaScript experience. Prior experience with GitHub Copilot varied; none had prior experience with Amazon Kiro. Prior to Phase 1, students were introduced to the tools to build baseline developer familiarity with AI-assisted tools and mitigate initial learning-curve effects. During this "dry-run" period, the developers built a small-scale website with alternative requirements to later phases. Project supervision was provided by industry partners, but the development of the artifacts was entirely implemented by the student-team to avoid confounding effects of more experienced industry developers.

### *Benchmark Application*

The benchmark was a full-stack web application comprising of user authentication, API call logging, and inventory management services, built with FastAPI, SQLite, HTML, CSS, and JavaScript. This architecture reflects a simplified but realistic multi-service production environment. Identical functional and non-functional requirements were used across all three phases.

### *Metrics*

Total development effort was measured in development hours. Time was self-reported at the feature level and cross-examined with tool logs when possible. Self-reporting is noted as an inherent limitation (Section VII).

Two AI-interaction efficiency metrics were tracked per feature: (1) prompt-to-feature ratio: the average number of prompts required to implement a feature; and (2) prompt success rate: the proportion of prompts producing usable output without significant rework, assessed subjectively by each developer in real time. A successful prompt was defined as one that produced output that was (a) directly usable without significant rework, or (b) meaningfully advanced feature implementation. This determination was binary (successful/unsuccessful) and made in real time during development. While subjective, this operationalization reflects the practical judgment developers apply during AI-assisted work and is acknowledged as a potential source of bias (Section VII).

Subjective cognitive load was assessed via administration of NASA-TLX administered to each developer at phase completion, measuring six dimensions: mental demand, physical demand, temporal demand, performance, and frustration. Given n=4, all values are descriptive means only.

At phase completion, an external industry evaluator, independent of the development team and informed of which application was tied to which phase of development, scored each application using a Requirements-to-Implementation Traceability Matrix (RITM) [9]. The results were not shared with the team until development across all three phases was complete. Eleven features were evaluated on a 0–2 scale (0 = not implemented; 1 = partially implemented; 2 = fully implemented); maximum score = 22 points. RITM measures requirement traceability only; it does not assess security, maintainability, or technical debt. The scoring rubric is provided in the Appendix.

### *Data Collection*

Structured data were recorded in a centralized spreadsheet per Kanban feature, capturing assigned developer, hours logged, prompts issued, successful prompts, and lines of code by category (AI-generated, human-written, human-corrected). Spreadsheet records were cross validated against GitHub commit history; discrepancies were resolved through commit diffs and session logs. GitHub Copilot Chat JSON exports and Amazon Kiro execution logs were processed through a custom Copilot Analysis Tool to compute derived metrics consistently across phases.

## V. RESULTS

All values are descriptive. NASA-TLX averages reflect n=4 responses per phase; no statistical significance testing was performed. Table I summarizes all key metrics.

*TABLE I. Summary of Key Metrics by Phase (NASA-TLX values are means, n=4)*

| Metric | Ph. 1 | Ph. 2 | Ph. 3 |
|---|---|---|---|
| **Development Hours** | 36.0 | 9.4 | 5.9 |
| **RITM Score (/ 22)** | 18 | 19 | 20 |
| **Prompt Success Rate** | 79.9% | 85.0% | 96.0% |
| **Prompts / Feature (mean)** | 2.66 | 1.92 | 2.42 |
| **Effort (NASA-TLX)** | 12.5 | 7.75 | 5.75 |
| **Mental Demand (NASA-TLX)** | 12.75 | 8.5 | 7.5 |
| **Temporal Demand (NASA-TLX)** | 8.25 | 6.75 | 6.0 |
| **Performance (NASA-TLX)** | 17.5 | 16.25 | 16.25 |
| **Frustration (NASA-TLX)** | 6.25 | 6.75 | 7.5 |
| **Physical Demand (NASA-TLX)** | 1.5 | 1.0 | 1.5 |

### *Development Efficiency (RQ1)*

Development hours declined monotonically: Phase 1 required 36.0 hours; Phase 2 required 9.4 hours (74% reduction); Phase 3 required 5.9 hours (a further 37% reduction; 84% total relative to Phase 1). This progression is directionally consistent with prior work demonstrating AI assistance can reduce task completion time by over 50% and extends those findings into AI-exclusive conditions [1, 2].

The additional reduction from Phase 2 to Phase 3 is directionally consistent with platform architecture contributing to efficiency differences, though the sequential design and accumulated developer familiarity prevent causal attribution (Section VII). Kiro's built-in specification-driven workflows may have reduced developer overhead in managing agent behavior.

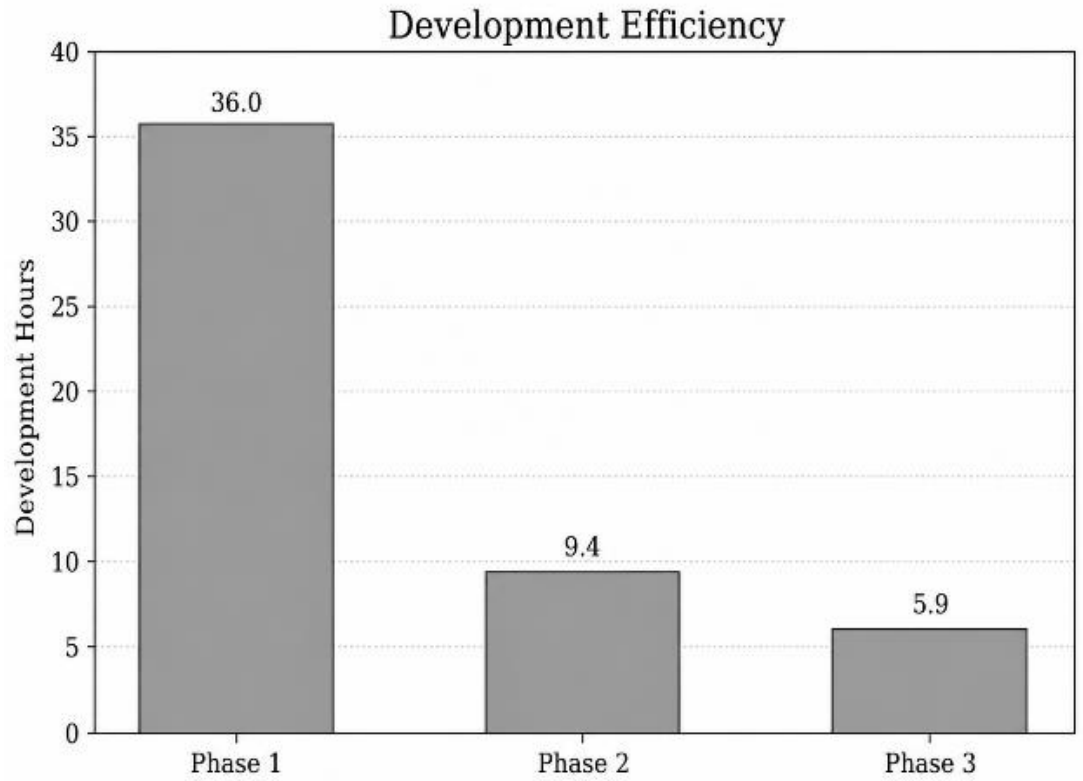

### *Code Quality (RQ2)*

RITM scores improved incrementally: Phase 1 = 18/22 (4 missing); Phase 2 = 19/22 (3 missing); Phase 3 = 20/22 (2 missing). Phase 3 achieved the highest requirement adherence despite the fewest development hours, suggesting efficiency and requirement fidelity need not trade off under AI-exclusive conditions. The AI-exclusive phases recorded higher requirement adherence than the partial agentic phase, contrary to the intuition that reduced human oversight should degrade this dimension of quality. RITM measures only requirement traceability; it does not assess security, maintainability, or technical debt. One plausible explanation is that higher AI autonomy frees developers to focus attention on requirement interpretation, prompt construction, and output validation, roles that may more directly influence final product quality than line-by-line code authorship. It is also possible that the team’s accumulated familiarity with the requirements across sequential phases contributed to quality improvements independent of tooling [10].

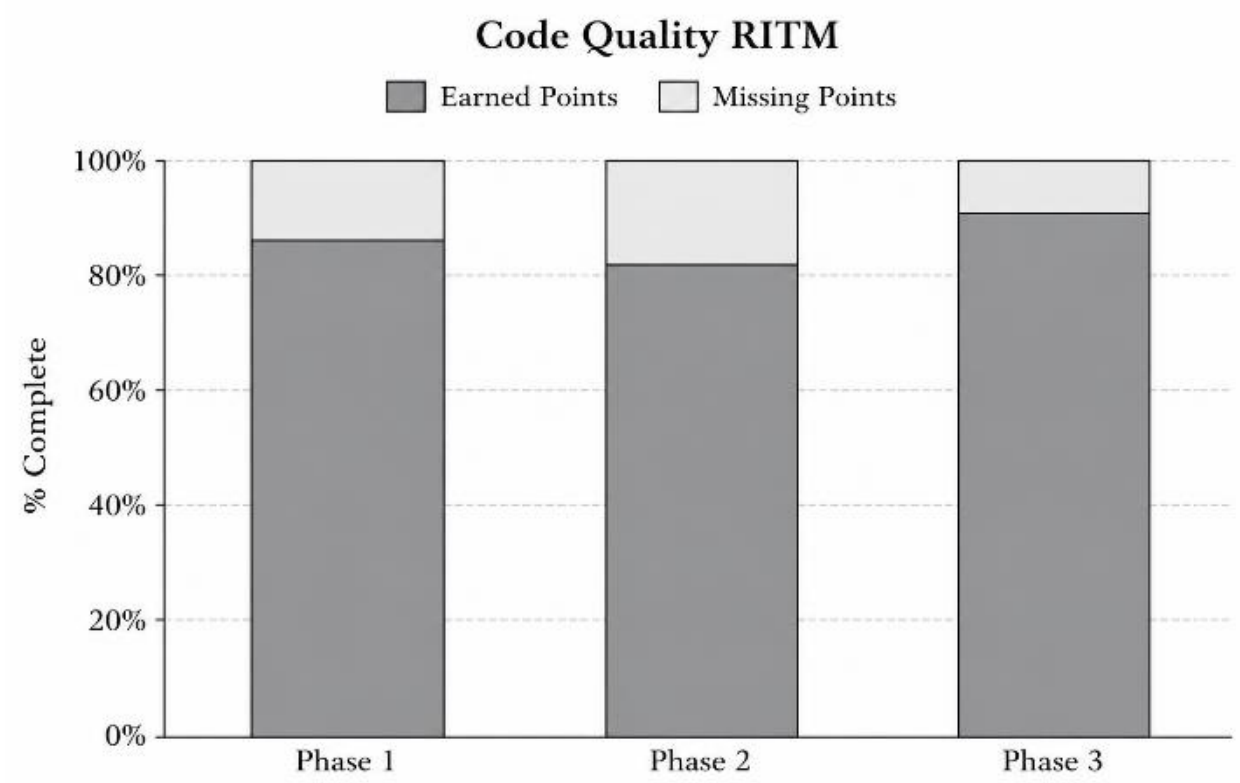

### *Developer Cognitive Load (RQ3)*

Effort declined from 12.5 (Phase 1) to 7.75 (Phase 2) to 5.75 (Phase 3), a 54% reduction. Mental Demand followed a similar trajectory: 12.75 (Phase 1) to 8.5 (Phase 2) to 7.5 (Phase 3), as did Temporal Demand 8.25 (Phase 1) to 6.75 (Phase 2) to 6.0 (Phase3). These reductions are directionally consistent with the recorded development hours and suggest productivity gains did not come at the cost of increased cognitive burden.

Frustration moved in the opposite direction: 6.25 (Phase 1), 6.75 (Phase 2), 7.5 (Phase 3). While absolute values remain low, this directional trend is consistent with the adoption friction developers may experience when transitioning from code authorship to agent orchestration. Phase 3 also introduced an unfamiliar tool environment.

Performance self-ratings were stable and high: 17.5, 16.25, 16.25, indicating that developers maintained a strong sense of contribution across all conditions, including AI-exclusive phases. Physical Demand was negligible and consistent (1.5, 1.0, 1.5), serving as an internal consistency check.

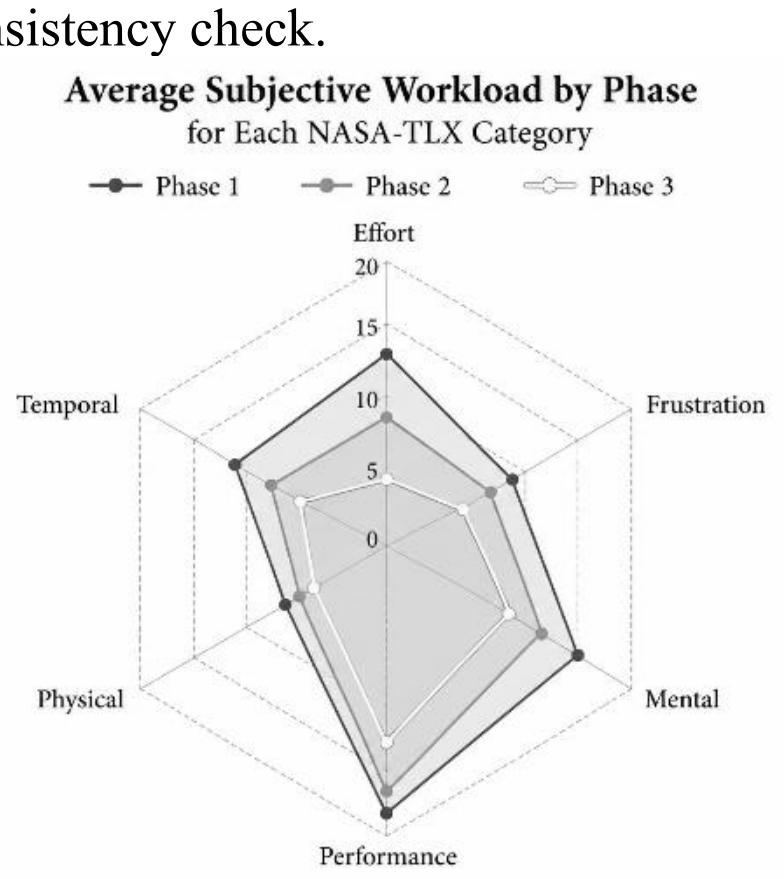

### *AI-Interaction Efficiency (RQ4)*

Prompt success rate increased consistently: 79.9% (Phase 1), 85.0% (Phase 2), 96.0% (Phase 3). This trend likely reflects both improving tool reliability and growing team proficiency in prompt formulation, factors that cannot be separated in the sequential design.

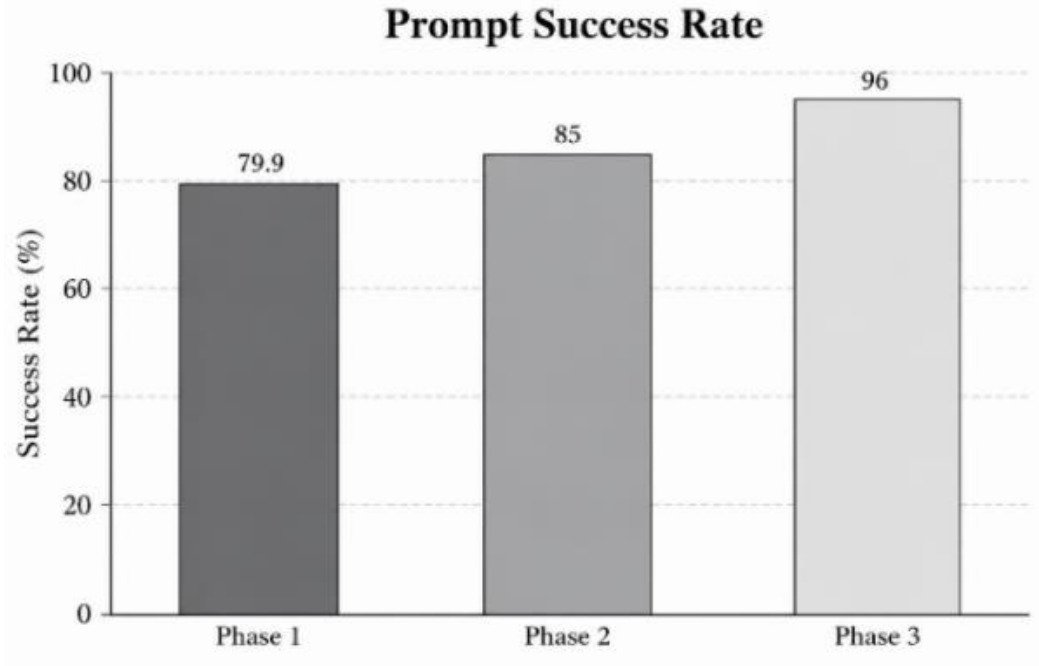

The prompt-to-feature ratio did not follow a monotonic trend: Phase 1 = 2.66, Phase 2 = 1.92 (lowest), Phase 3 = 2.42. This divergence is consistent with a platform-level architectural difference: GitHub Copilot favored fewer, broader exchanges, while Kiro's specification-driven workflow decomposes features more granularly, requiring more prompts per feature while achieving higher per-prompt success. Total prompt volume is an incomplete proxy for interaction efficiency.

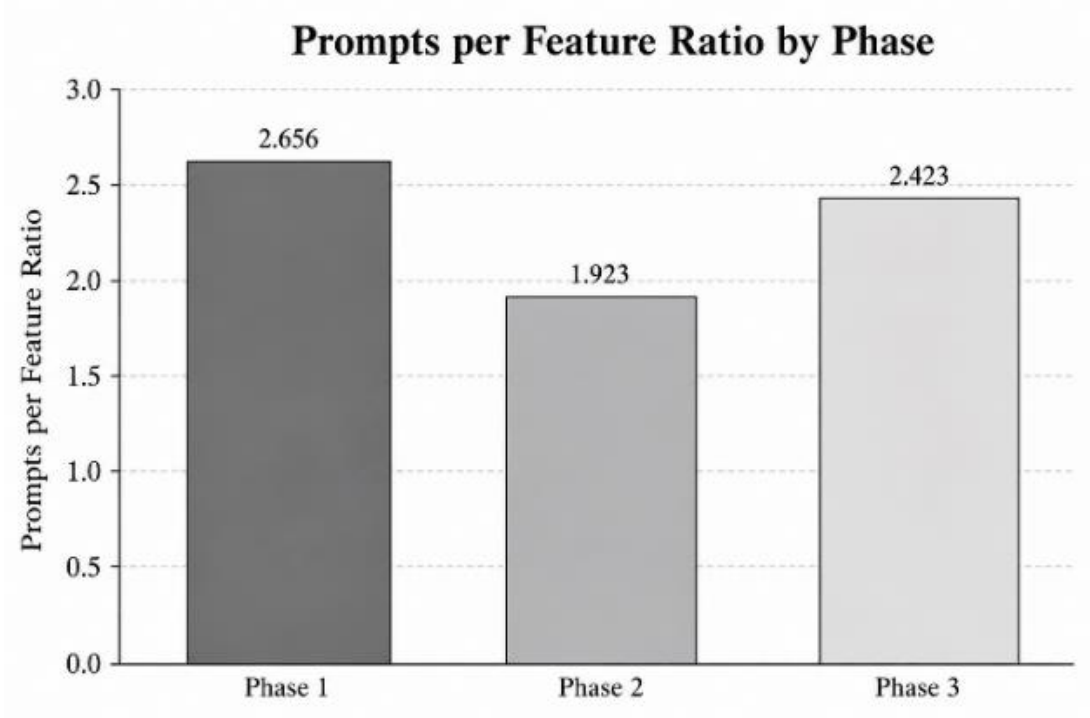

## VI. DISCUSSION

### *AI Autonomy, Productivity, and Quality Tradeoffs*

The 84% total reduction in development hours is directionally consistent with prior controlled experiments [1, 2]. The incremental RITM improvement provides preliminary evidence that reduced human oversight of code authorship was not associated with degraded requirement adherence in this well-scoped greenfield setting. One interpretation is that agentic development shifts developer attention toward higher-leverage activities, requirement interpretation, prompt design, and output validation, that may directly influence requirement fidelity. This interpretation must be treated cautiously given the sequential confound [10]. Additionally, RITM reflects only requirement traceability; other quality dimensions were not assessed.

The additional 37% developer hours reduction from Phase 2 to Phase 3, at nominally equivalent autonomy, is directionally consistent with tooling architecture mattering independently of AI autonomy level. This suggests platform selection may warrant empirical evaluation rather than commodity treatment.

### *Human-AI Orchestration Patterns*

GitHub Copilot's design favored fewer, broader prompt exchanges; Amazon Kiro's specification-driven workflow decomposed tasks more granularly. Neither pattern is inherently superior: Kiro required more prompts per feature but achieved a substantially higher per-prompt success rate (96.0% vs. 85.0%) and superior RITM scores in fewer man-days. Evaluating platforms on a single efficiency metric is likely to be misleading; organizations should examine prompt volume, success rate, development hours, and quality measures together.

The rising prompt success rates across phases are plausibly attributable to both tool quality and growing team skill in prompt formulation. In agentic environments, prompt engineering and requirement translation become critical competencies [4].

### *Cognitive and Organizational Implications*

The pattern of declining effort alongside rising frustration is practically significant. Agentic development reduces the mechanical burden of development while introducing new friction around tool learning, reduced code authorship, and loss of direct agency over the codebase. The stable performance self-ratings even across AI-exclusive workflows are an encouraging counterpoint, suggesting the shift from writing code to agent orchestration need not undermine developer engagement if the transition is managed thoughtfully. For organizations deploying agentic workflows, these findings point to the importance of structured onboarding, change management, and role redesign.

### *Implications for Practice*

Several directional implications emerge. First, fully agentic development may not degrade requirement adherence relative to partial assistance in well-scoped greenfield settings. Second, platform selection appears to matter independently of AI autonomy level. Third, rising frustration under fully agentic conditions points to a human factors dimension warranting dedicated attention. These implications should be generalized only with caution; production environments typically involve legacy codebases, evolving requirements, security constraints, and distributed teams.

## VII. LIMITATIONS

### *Sequential Design and Learning Effects*

The most significant limitation is the sequential phase design. The same four-person team implemented the same application across all three phases, accumulating familiarity with the architecture, requirements, prompting strategies, and agentic workflows over time. Improvements in later phases may reflect learning effects in addition to tooling or autonomy differences. Because phases were neither randomized nor counterbalanced, treatment effects cannot be cleanly separated from accumulated developer experience. This limitation is especially relevant for the Phase 2 vs. Phase 3 comparison.

### *Sample Size and Generalizability*

This study involved a single team of four student developers on a small greenfield application. Findings may not generalize to professional engineering organizations, larger teams, legacy systems, or projects with complex compliance constraints. NASA-TLX aggregates based on n=4 responses are sensitive to individual outlier responses; no statistical significance testing was performed. The student-developer context introduces an experience profile that may differ substantially from how experienced software engineers interact with agentic AI systems in enterprise settings. Replication across multiple teams, organizations, and project types is necessary before broader conclusions can be drawn.

### *Measurement and Construct Limitations*

Time tracking was self-reported rather than system validated. Prompt success rate was assessed without standardized rubric, introducing potential inconsistency. NASA-TLX scores reflect subjective perceptions collected at phase conclusion and may be influenced by overall project sentiment. RITM measures requirement traceability only; improvements should not be interpreted as evidence of comprehensive software quality improvement. Because the external industry evaluator was informed of phase identity, complete evaluator blinding cannot be guaranteed.

### *Tooling and Reproducibility Constraints*

The findings reflect specific behavior of GitHub Copilot and Amazon Kiro during the study period. AI tools evolve rapidly and these results may not generalize to future versions. Exact model versions, runtime configurations, agent permissions, and session parameters were not systematically archived, limiting full procedural reproducibility.

### *Scope and Duration*

Each phase involved short development cycles on a constrained benchmark. The study cannot assess technical debt accumulation, maintainability degradation, onboarding complexity, developer skill atrophy, or sustained organizational adaptation to agentic workflows.

## VIII. FUTURE WORK

### *Expanded Design and Participant Diversity*

Future research should replicate this study across multiple independent teams, organizational settings, and developer experience levels. Randomized or parallelized experimental designs would substantially strengthen causal interpretation by separating learning effects from treatment effects [10].

### *Expanded Quality and Longitudinal Evaluation*

Future work should incorporate broader objective quality metrics: static analysis results, cyclomatic complexity, security vulnerabilities, test coverage, defect rates, and maintainability indicators. Longer-duration studies should examine technical debt accumulation, refactoring effort, onboarding complexity, and bug-fix rates over extended development cycles.

***Broader Platform Comparisons***

Future comparative studies should evaluate a broader range of AI development environments, orchestration models, and tooling architectures under controlled conditions [11], paired with the randomized designs described above. Such work would provide more actionable guidance for platform selection and workflow integration.

## IX. CONCLUSION

This paper presents an exploratory pilot study examining how increasing levels of AI autonomy and differing agentic platforms affect software development productivity, code quality, and developer cognitive load. Across three controlled phases, developer hours declined by 87%, RITM scores improved incrementally, and self-reported effort and mental demand decreased substantially. Amazon Kiro (Phase 3) was associated with lower man-days, higher RITM scores, and higher prompt success rates than GitHub Copilot (Phase 2) at equivalent AI autonomy levels, providing directional evidence that platform architecture may be a relevant variable. Prompts per feature and frustration did not follow this pattern.

The dimension that moved against the prevailing trend, rising developer frustration, highlights a human factors cost of agentic adoption that warrants dedicated investigation and careful attention in organizational deployment strategies.

These findings must be interpreted within significant constraints: a single four-person student team, a small greenfield application, non-randomized sequential phases, and a sample size that precludes statistical inference. This work is offered as a methodological foundation and source of directional hypotheses for future, more rigorous investigations. The evaluation framework employed here is adaptable for future comparative studies, with the understanding that full procedural reproducibility is limited by unarchived runtime configurations.

## ACKNOWLEDGMENTS

This study was conducted in partnership with Clemson University and industry partners as part of Clemson's School of Computing Industry Capstone Program, whose support enabled access to development resources and tooling across all three experimental phases. The authors declare no financial conflict of interest with any AI tool vendor studied. The views expressed are those of the authors and do not represent the views of industry partners or Clemson University.

## REFERENCES.

## APPENDIX
## RITM Scoring Rubric

Each of the eleven evaluated features was scored on a 0–2 scale by an independent evaluator not informed of which phase produced each implementation. The rubric below defines each score level with representative examples. Scores of 1 were assigned when a feature demonstrated meaningful partial progress but had one or more critical functional gaps.

*TABLE A1. RITM Feature Scoring Rubric*

| Score | Definition | Representative Example |
|---|---|---|
| **0** | Feature absent or completely non-functional. No meaningful implementation exists. | Login page is blank; no authentication routes are defined. |
| **1** | Feature partially implemented. Core structure exists but critical functionality is missing, broken, or incomplete. | Login form renders and accepts input but does not authenticate against the database; session is not created. |
| **2** | Feature fully implemented. All specified requirements are met and the feature functions as intended. | Login form authenticates against the database, creates a session, redirects authenticated users, and rejects invalid credentials with appropriate error messages. |